%% file: main.tex
\documentclass[conference]{IEEEtran}
\IEEEoverridecommandlockouts

\usepackage{graphicx}
\usepackage{textcomp}
\usepackage{xcolor}
\usepackage[letterpaper, total={180mm,229mm}]{geometry}
\def\BibTeX{{\rm B\kern-.05em{\sc i\kern-.025em b}\kern-.08em
    T\kern-.1667em\lower.7ex\hbox{E}\kern-.125emX}}

\usepackage{graphicx}
\usepackage{amsmath,amssymb,amsfonts}
\usepackage[noend,linesnumbered,ruled,vlined]{algorithm2e}
\usepackage{booktabs}
\usepackage{listings}
\usepackage{microtype}
\usepackage{subcaption}
\usepackage{multirow}
\usepackage[outline]{contour}
\contourlength{0.7pt}

\usepackage{amsmath}
\usepackage{amsthm}
\usepackage{hyperref}
\usepackage{tabularx}
\usepackage{pgfplots}
\usepackage[edges]{forest}
\forestset{
  dir tree/.style={
    for tree={
      parent anchor=south west,
      child anchor=west,
      anchor=mid west,
      inner ysep=-1pt,
      grow'=0,
      align=left,
      edge path={
        \noexpand\path [draw, \forestoption{edge}] (!u.parent anchor) ++(0.5em,-0.5em) |- (.child anchor)\forestoption{edge label};
      },
      font=\footnotesize,
      if n children=0{}{
        delay={
          prepend={[,phantom, calign with current]}
        }
      },
      fit=band,
      before computing xy={
        l=2em
      }
    },
  }
}

\renewcommand{\footnoterule}{%
	\kern -3pt
	\hrule
	\kern 2pt
}

\usepackage{microtype}
\usepackage{rotating}
\SetCommentSty{mycommfont}
\usepackage[sorting=none, mincitenames=1,backend=bibtex,doi=false,isbn=false,url=false]{biblatex} %
\usepackage{scalerel}
\usepackage{tikz} 

\usetikzlibrary{svg.path}

\definecolor{orcidlogocol}{HTML}{A6CE39}
\tikzset{
  orcidlogo/.pic={
    \fill[orcidlogocol] svg{M256,128c0,70.7-57.3,128-128,128C57.3,256,0,198.7,0,128C0,57.3,57.3,0,128,0C198.7,0,256,57.3,256,128z};
    \fill[white] svg{M86.3,186.2H70.9V79.1h15.4v48.4V186.2z}
                 svg{M108.9,79.1h41.6c39.6,0,57,28.3,57,53.6c0,27.5-21.5,53.6-56.8,53.6h-41.8V79.1z M124.3,172.4h24.5c34.9,0,42.9-26.5,42.9-39.7c0-21.5-13.7-39.7-43.7-39.7h-23.7V172.4z}
                 svg{M88.7,56.8c0,5.5-4.5,10.1-10.1,10.1c-5.6,0-10.1-4.6-10.1-10.1c0-5.6,4.5-10.1,10.1-10.1C84.2,46.7,88.7,51.3,88.7,56.8z};
  }
}

\newcommand\orcidicon[1]{\href{https://orcid.org/#1}{\mbox{\scalerel*{
\begin{tikzpicture}[yscale=-1,transform shape]
\pic{orcidlogo};
\end{tikzpicture}
}{|}}}}

\begin{document}

\onecolumn
\thispagestyle{empty}
\twocolumn
\setcounter{page}{1}
\setcounter{figure}{0}

\title{BiKA: Kolmogorov-Arnold-Network-inspired Ultra Lightweight Neural Network Hardware Accelerator
}

\author{
\IEEEauthorblockN{Yuhao Liu$^{1,2,3}$ \orcidicon{0000-0002-7281-2126}, \textit{Student Member, IEEE}, Salim Ullah$^{1}$ \orcidicon{0000-0002-9774-9522}, Akash Kumar$^{1}$ \orcidicon{0000-0001-7125-1737}, \textit{Senior Member, IEEE}}
\IEEEauthorblockA{$^{1}$Ruhr University Bochum, Germany $^{2}$Dresden University of Technology, Germany\\
$^{3}$Center for Scalable Data Analytics and Artificial Intelligence (ScaDS.AI Dresden/Leipzig), Germany\\
Email: \{yuhao.liu, salim.ullah, akash.kumar\}@rub.de}
}

\maketitle
\begin{abstract}
		 \input{ISQED2026/abstract}

\end{abstract}%

\thispagestyle{empty}


\section{Introduction}
\label{introduction}
\input{ISQED2026/introduction}

\section{Implementation}
\label{imp}
\input{ISQED2026/implementation}

\section{Evaluation}
\label{evaluation}
\input{ISQED2026/evaluation}

\section{Conclusion and Further Works}
\label{concl}
\input{ISQED2026/conclusion}

\section*{Acknowledgments}
This research was supported in part by the Deutsche Forschungsgemeinschaft (DFG) under the X-ReAp project (Project number 380524764) and by the Center for Scalable Data Analytics and Artificial Intelligence (ScaDS.AI Dresden/Leipzig), Germany.

%
%
\renewcommand{\bibfont}{\footnotesize}
\printbibliography
\end{document}

%% file: ISQED2026/abstract.tex
Lightweight neural network accelerators are essential for edge devices with limited resources and power constraints. While quantization and binarization can efficiently reduce hardware cost, they still rely on the conventional \emph{Artificial Neural Network} (ANN) computation pattern. The recently proposed \emph{Kolmogorov-Arnold Network} (KAN) presents a novel network paradigm built on learnable nonlinear functions. However, it is computationally expensive for hardware deployment. Inspired by KAN, we propose BiKA, a multiply-free architecture that replaces nonlinear functions with binary, learnable thresholds, introducing an extremely lightweight computational pattern that requires only comparators and accumulators. Our FPGA prototype on \emph{Ultra96-V2} shows that BiKA reduces hardware resource usage by $27.73\%$ and $51.54\%$ compared with binarized and quantized neural network systolic array accelerators, while maintaining competitive accuracy. BiKA provides a promising direction for hardware-friendly neural network design on edge devices.

%% file: ISQED2026/introduction.tex
With the rapidly increasing complexity of modern \emph{Neural Networks} (NNs), reducing the hardware cost of NN accelerators has become a key research topic. Related researchers have thoroughly explored a wide range of techniques to simplify computation and reduce memory consumption, including quantization, binarization, and approximate arithmetic. For instance, ~\emph{FINN}~\cite{Umuroglu2017, Blott2018},~\emph{HLS4ML}~\cite{HLS4ML, hls4ml2}, and~\emph{LogicNets}~\cite{Umuroglu2020} proposed the efficiency of \emph{Binarized Neural Network} (BNNs) and \emph{Quantized Neural Network} (QNNs) on FPGAs. Other works have explored approximate multipliers, such as Ullah et al.~\cite{approx1, approxNN}, or alternative neural paradigms, including \emph{Spiking Neural Networks} (SNNs), which are implemented in systems like \emph{SpiNNaker2}~\cite{SpiNNaker2} and \emph{S2N2}~\cite{Khodamoradi2021}. However, most existing accelerator research is focusing on conventional \emph{Artificial Neural Networks} (ANNs). This motivates the exploration of emerging network models that provide new paradigms of trade-offs in accuracy, resource consumption, and hardware friendliness. 

The \emph{Kolmogorov-Arnold Network} (KAN)~\cite{kan} is a recently introduced model that replaces the multiplication and activation operations in conventional ANNs with learnable nonlinear functions, offering a fundamentally different computational paradigm and new opportunities for hardware-efficient design. Despite its promising potential, recent research still lacks the exploration of KAN-based hardware accelerators, especially for resource-limited edge devices.

\begin{figure}[t]
    \centering
    \begin{minipage}{1\columnwidth}
        \centerline{\includegraphics[width=\columnwidth]{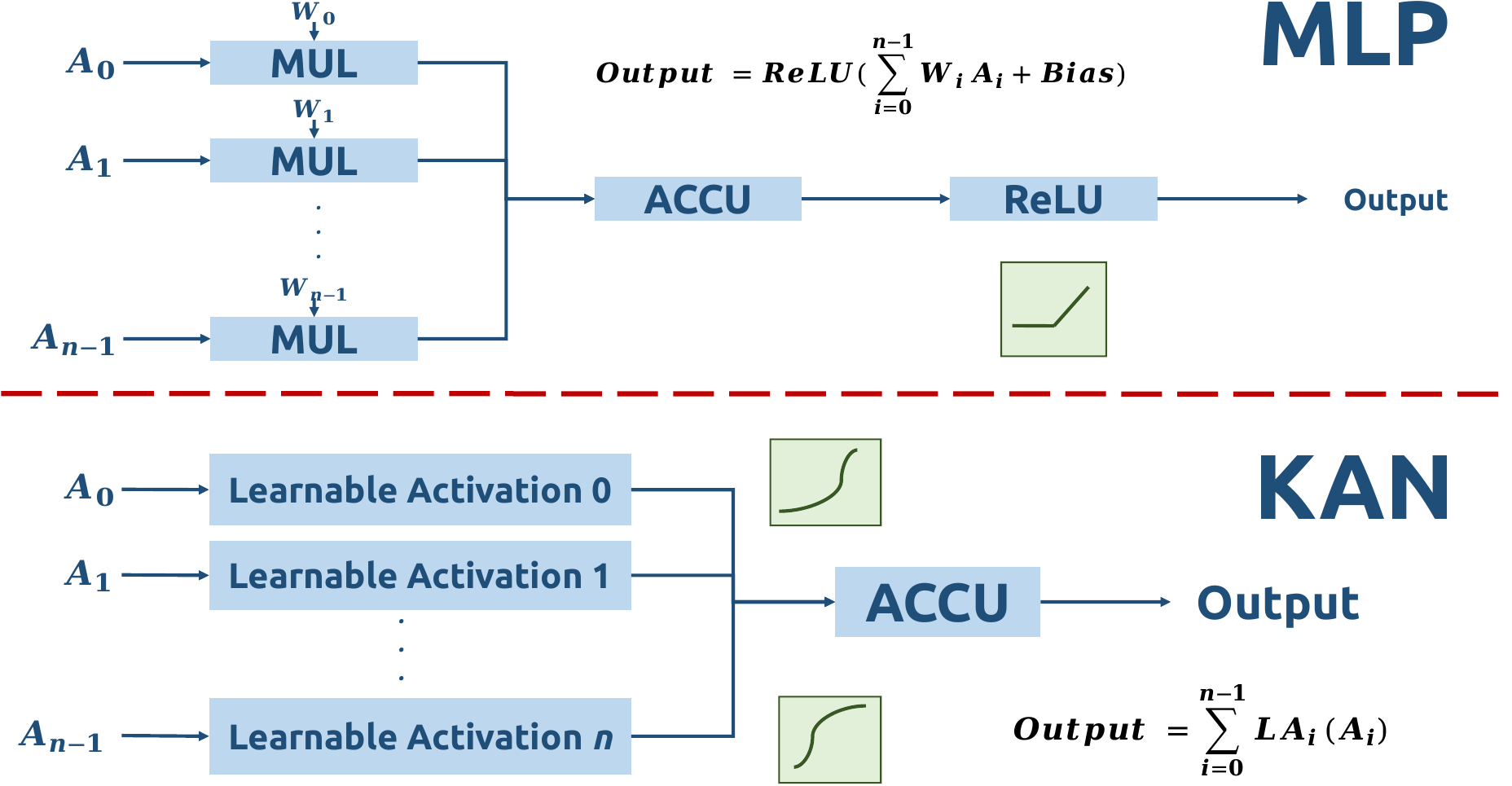}}
        \vspace{10pt}
        \caption{Difference between the computation of single neuron in MLP (up) and KAN (down)}
        \label{KANvsMLP}
    \end{minipage}
\end{figure}

\subsection{Background}

\subsubsection{What is Kolmogorov-Arnold Network}
From the 1960s, the Kolmogorov-Arnold representation theorem and related works~\cite{kolmogorov1957representation, arnold2009representation, braun2009constructive} proved that every continuous function can be represented as a sum of functions of fewer variables. Building upon this principle, Liu et al.~\cite{kan} introduced the Kolmogorov–Arnold Network (KAN) in 2024 as a new neural architecture aimed at improving interpretability and reducing model complexity. As shown in~\autoref{KANvsMLP}, KAN replaces the \emph{Multiplication-and-Activation} (MAC) computations in a \emph{Multilayer Perceptron} (MLP) with a set of learnable nonlinear functions $LA_i(x)$ applied to each input activation $A_i$. The outputs of these functions are summed as the neuron output. Previous work~\cite{qkan} evaluated the accuracy of KAN against MLP on MNIST~\cite{deng2012mnist}. KAN shows a competitive accuracy ($98.52\%$) compared to MLP ($98.02\%$).

Despite its potential, KAN introduces several challenges for both training and 
hardware acceleration:

\begin{itemize}
    \item \emph{Difficulties in Training}: Compared to the weight and bias in ANN, learnable nonlinear functions in KAN are more complex and consume more computational resources in training. For example, based on our experiment, training even a small three-layer KAN model (64/32/10 neurons) using the native \emph{pykan} library~\cite{kan} can exceed 48\,GB of memory on an NVIDIA A40 GPU.
    \item \emph{Difficulties in Accelerator Design}: Learnable nonlinear function design is a novel computational pattern introduced by KAN. However, nonlinear functions complicate hardware implementation and cannot be directly applied to common simplification techniques, such as quantization, because KAN is both multiply-free and weight-free. Therefore, it needs a novel design paradigm for lightweight hardware accelerators on the edge. 
\end{itemize}

\begin{table}
    \centering
    \caption{Resource Consumption of MLP and KAN on FPGA Platform Based on Post-implementation in Previous Work}
    \resizebox{1\columnwidth}{!}
    {
        \begin{tabular}{cccccccc}
            \toprule
            Dataset                   & Model Type & Model Size    & Frequency               & LUT     & FF      & DSP   & BRAM \\ \midrule
            \multirow{2}{*}{Wine}     & MLP        & 13,32,8,3     &                         & 6974    & 9936    & 17    & 2    \\
                                      & KAN        & 13,4,3        &                         & 146843  & 74741   & 950   & 132  \\
            \multirow{2}{*}{Dry Bean} & MLP        & 16,20,15,10,7 &                         & 8894    & 11328   & 17    & 0    \\
                                      & KAN        & 16,2,7        &                         & 1677558 & 734544  & 9111  & 781  \\
            \multirow{2}{*}{Mashroom} & MLP        & 8,64,64,2     &                         & 10932   & 18903   & 17    & 4    \\
                                      & KAN        & 8,24,2        &                         & 3112275 & 1337291 & 16299 & 1347 \\ \bottomrule
        \end{tabular}
    }
    \label{previous_kan_hardware_tab}
\end{table}


\subsubsection{Previous Accelerators Design for KAN} 

As an emerging network structure, several recent works explored the design of a hardware accelerator for KAN. For instance, Tran et al.~\cite{tran2024KAN} first explored the hardware accelerator design of KAN on an FPGA based on a handmade \emph{High-Level Synthesis} (HLS) implementation, comparing it with MLP. Huang et al.~\cite{KAN-Mixed-signal} first presented the KAN accelerator based on Analog-digital-mixed hardware produced with TSMC 22 nm node. Yin et al.~\cite{qkan} introduced their quantized KAN accelerator based on Look-Up Table (LUT) optimization for FPGA platforms. 

As shown in \autoref{previous_kan_hardware_tab}, Tran et al.~\cite{tran2024KAN} implemented four small KAN-based MLP classifiers using \emph{Vitis HLS}. Their results show that directly mapping native KAN to hardware leads to extremely high resource usage (e.g., ~3.1M LUTs for only 34 fully connected
kernels), even for toy-scale models. These findings highlight a fundamental challenge: the nonlinear-function-based computation in KAN is prohibitively expensive for FPGA-based or edge-oriented accelerator designs.

Huang et al.~\cite{KAN-Mixed-signal} explored the hardware-software co-design framework for the lightweight edge accelerator based on N:1 Time Modulation Dynamic Voltage input generator for RRAM-ACIM. However, the proposed analog-digital signal mixed hardware in this work is complicated and not easily scalable in design and manufacturing, and cannot be simply integrated into conventional FPGA/ASIC design flows for general edge devices.

Yin et al.~\cite{qkan} applied global and fine-grained post-training quantization to KAN, converting each learned nonlinear function into a low-bit lookup table. This enables KAN inference using integer arithmetic and reduces computational cost. However, their method generates a model-specific hardware design: each trained network requires regenerating all LUTs, rebuilding the 
hardware architecture, and performing full synthesis and bitstream generation, which currently takes 6–10 hours. Moreover, because their accelerator implements the entire network structure on the FPGA, resource consumption grows accordingly with the model size, because every nonlinear function requires one LUT. For example, a three-layer KAN with 64/32/10 neurons for MNIST, as presented in their work, requires a maximum of 65,680 LUT-6s. Applying their approach to the VGG-like convolutional network used in our BiKA evaluation for the CIFAR-10~\cite{Krizhevsky2009} dataset would require approximately 4.3 million LUT-6s, which exceeds the capacity of typical edge FPGAs.

\begin{figure}[t]
    \centering
    \begin{minipage}{1\columnwidth}
        \centerline{\includegraphics[width=\columnwidth]{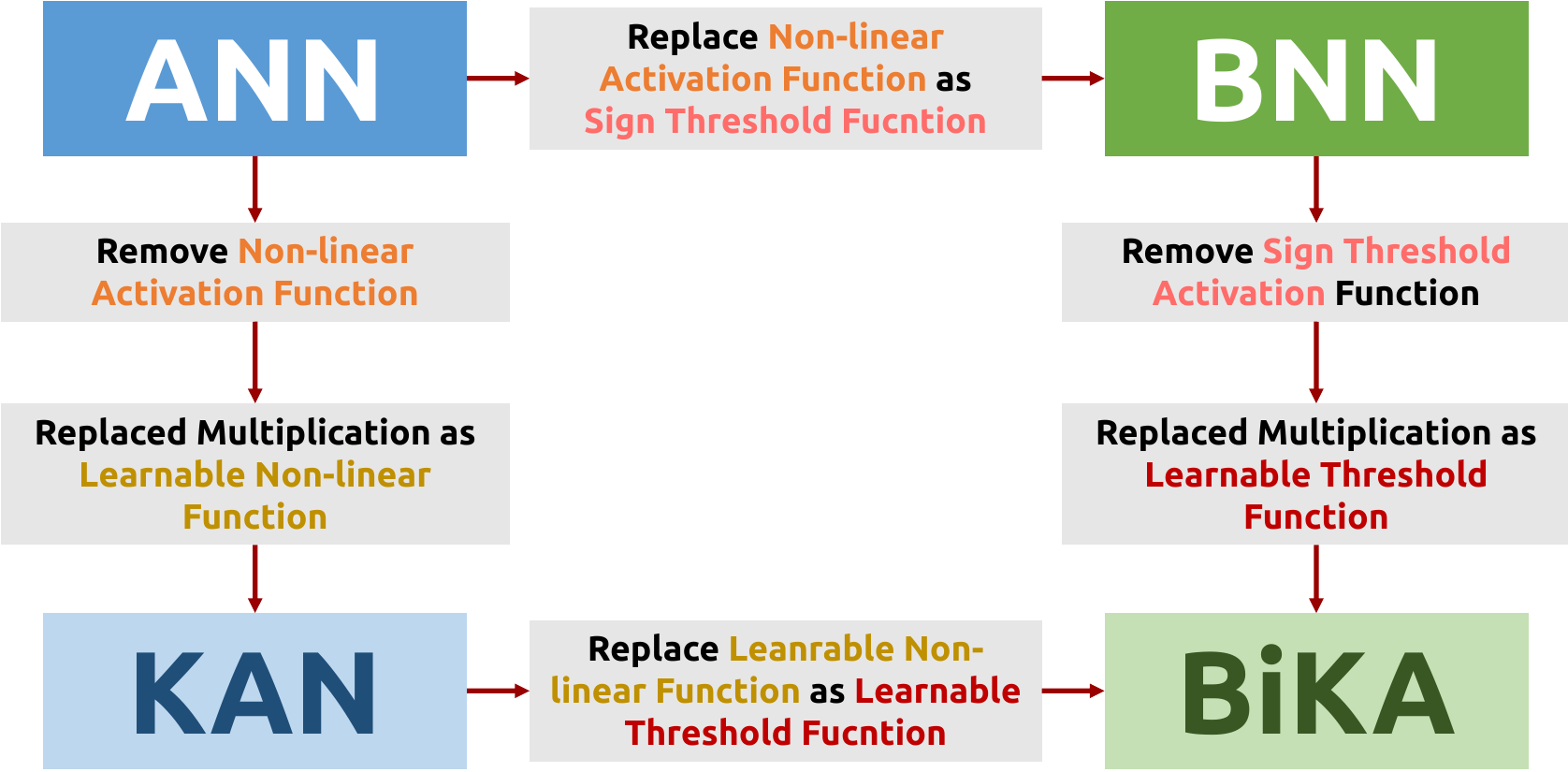}}
        \vspace{10pt}
    \end{minipage}
        \caption{Difference between the ANN, BNN, KAN, and BiKA}
        \label{KANvsANNvsBNNvsBiKA}
\end{figure}

\subsection{Motivation}

The prior works discussed above suggest that the fundamental challenge of deploying KAN on edge-oriented hardware accelerators lies in implementing its learnable nonlinear activation unit. Efficiently simplifying or replacing this operator is essential for enabling a KAN-based accelerator on FPGA/ASIC realization. Moreover, maintaining compatibility with conventional systolic-array architectures would significantly improve scalability to larger network models.

Therefore, we thoughtfully rethink the conversion between ANN, KAN, and QNN, especially the 1-bit quantized BNN. As shown in~\autoref{KANvsANNvsBNNvsBiKA},  typical BNNs replace the fixed nonlinear activation in ANN with a fixed \emph{Sign} threshold function. As a mirror design, we consider replacing the learnable nonlinear functions in KAN with learnable threshold functions to propose a \emph{Binarized KAN} (BiKA). From another perspective, since removing the fixed nonlinear activation function and replacing the multiplication with learnable nonlinear functions can convert the ANN to KAN, correspondingly, removing the fixed threshold function and replacing the multiplication with learnable threshold functions should suggest the existence of a potential binarized KAN structure, BiKA, converting from the binarized ANN. Therefore, based on this BiKA network, we can expect a simplified, ultra-lightweight model structure consisting of only comparators and integer adders, supporting the systolic-array structure, without any multiplication or nonlinear operations. Furthermore, considering BiKA is a fully integer-based network, it does not aim to preserve the interpretability properties of KAN. Instead, it follows a different design goal: achieving extremely low-cost computation for efficient hardware acceleration.

\subsection{Contributions}

\begin{itemize}

    \item Extending on our prior abstract in~\cite{bika_poster}, we propose BiKA, a new KAN-inspired neural network that replaces KAN’s expensive learnable nonlinear functions with binary learnable thresholds. This design removes all multiplications and nonlinear processing, resulting in an extremely lightweight computation model based only on comparators and integer adders.

    \item We provide a mathematical explanation showing that a KAN nonlinear function can be approximated using a learnable threshold. This gives a clear and direct theoretical basis for the BiKA architecture.

    \item We develop a \emph{PyTorch}~\cite{Paszke2019} and CUDA training framework for BiKA, and evaluate both MLP and CNN versions of the model on MNIST and CIFAR-10. We further analyze BiKA’s training behavior and explain the source of its accuracy loss. \footnote{An open-source reference implementation and training library of BiKA is available at \url{https://github.com/liuyh-Horizon/BiKA}. A mirrored repository for archival purposes is maintained at \url{https://git.noc.ruhr-uni-bochum.de/liuyuhyc/bika}.
.}

    \item We design a systolic-array accelerator for BiKA and compare it with BNN and QNN systolic-array accelerators. BiKA fits well into this structure because its operations are fully integer and multiplication-free.

    \item FPGA implementations on an Ultra96-V2 show that the BiKA accelerator reduces hardware resource usage by $27.73\%$ compared with BNN and $51.54\%$ compared with QNN, while keeping $<1\%$ accuracy loss on MNIST and about $10\%$ on CIFAR-10.  These results show that BiKA is a potential and highly efficient option for ultra-lightweight neural network acceleration on edge devices.
    
\end{itemize}

\subsection{Organization}

This manuscript is structured as follows: Section I discusses the background of KAN and previous KAN accelerator designs, presenting our motivation and contribution introduced in this work. Section II introduces the mathematical principle of BiKA and the design of BiKA accelerators. Section III presents the training results and hardware resource consumption of BiKA and its accelerator in comparison with QNN and BNN. Section IV discusses the further potential improvement and concludes the contents of this paper.

%% file: ISQED2026/implementation.tex
\subsection{Approximate the Nonlinear Function as the Form of Thresholds}
Considering the complexity of a nonlinear function, if we want to approximate it to a single threshold function in BiKA, we need to prove that it can be converted into a format constructed with a threshold-based representation first, for further approximation. Because the modern digital system cannot process continuous functions in principle, we can use a discrete function $f(x)$ as shown in~\autoref{piecewise_constant} to approximate the continuous nonlinear function in a certain input range [$s_0$, $s_t$), which can also be considered as one piecewise constant function with $t$ input slots, $s_i\leqslant x<  s_{i+1}$ $(0\leqslant i\leqslant t-1)$. When $t$ is large enough, which means that each input slot, $s_i\leqslant x<  s_{i+1}$, is small enough, $f(x)$ can approximately represent any continuous nonlinear function. 

\begin{equation}
   \small
   \begin{aligned}
       f(x)=   \begin{cases}
                   O_0 & \text{ if } s_0\leqslant x <  s_1 \\
                   O_1 & \text{ if } s_1\leqslant x < s_2 \\
                   \cdots \\
                   O_{t-1} & \text{ if } s_{t-1}\leqslant x < s_{t}
               \end{cases}
   \end{aligned}
   \label{piecewise_constant}
\end{equation}

Therefore, as shown in~\autoref{weightThres}, we define $t$ weighted threshold activations, $\alpha_iThres_{i}(x)$ $(0\leqslant i\leqslant t-1)$, based on input slots, $s_i\leqslant x< s_{i+1}$ $(0\leqslant i\leqslant t-1)$, where $\alpha_i$ is the weight for each threshold and the left end $s_i$ of input slot is the threshold value for each $Thres_{i}(x)$. When input $x$ is smaller than threshold value $s_i$, $Thres_{i}(x)$ outputs $-1$. Else, $Thres_{i}(x)$ outputs $1$.

\begin{equation}
    \small
    \begin{aligned}
        & \alpha_iThres_{i}(x)=  \begin{cases}
                        \alpha_i &  s_i\leqslant x \\
                        -\alpha_i &  x <  s_{i}
                    \end{cases} \\
    \end{aligned}
    \label{weightThres}
\end{equation}

\begin{equation}
    \small
    \begin{aligned}
        Assume \ \ f(x) \approx f'(x)=\sum_{i=0}^{t-1}\alpha_iThres_{i}(x)
    \end{aligned}
    \label{constructedF}
\end{equation}

Based on the sum of all weighted threshold activations, we construct one function, $f'(x)$, as shown in~\autoref{constructedF}, attempting to approximate the original discrete nonlinear function as $f(x)\approx f'(x)$.  

\begin{equation}
    \small
    \begin{matrix}
        \begin{aligned}
            & When\ s_i\leqslant x<s_{i+1}\ ,\\
            & \begin{cases}
                \alpha_0Thres_{0}(x) = \alpha_0 \\
                \alpha_1Thres_{1}(x) = \alpha_1 \\
                \cdots \\
                \alpha_iThres_{i}(x) = \alpha_i \\
                \cdots \\
                \alpha_{t-2}Thres_{t-2}(x) = -\alpha_{t-2} \\
                \alpha_{t-1}Thres_{t-1}(x) = -\alpha_{t-1}
              \end{cases} \\
        \end{aligned}
     &  
     \begin{aligned}
        & Therefore\ ,\\
        & \begin{aligned}
            f(x)& = O_i \approx f'(x) \\ 
                & =\sum_{i=0}^{t-1}\alpha_iThres_{i}(x) \\
                & =\sum_{l=0}^{i}\alpha_l-\sum_{r=i+1}^{t-1}\alpha_r 
        \end{aligned} \\
    \end{aligned} \\
    \end{matrix}
    \label{Oiandweights}
\end{equation}

\begin{figure}[t]
    \centering
    \begin{minipage}{1\columnwidth}
        \centerline{\includegraphics[width=0.9\columnwidth]{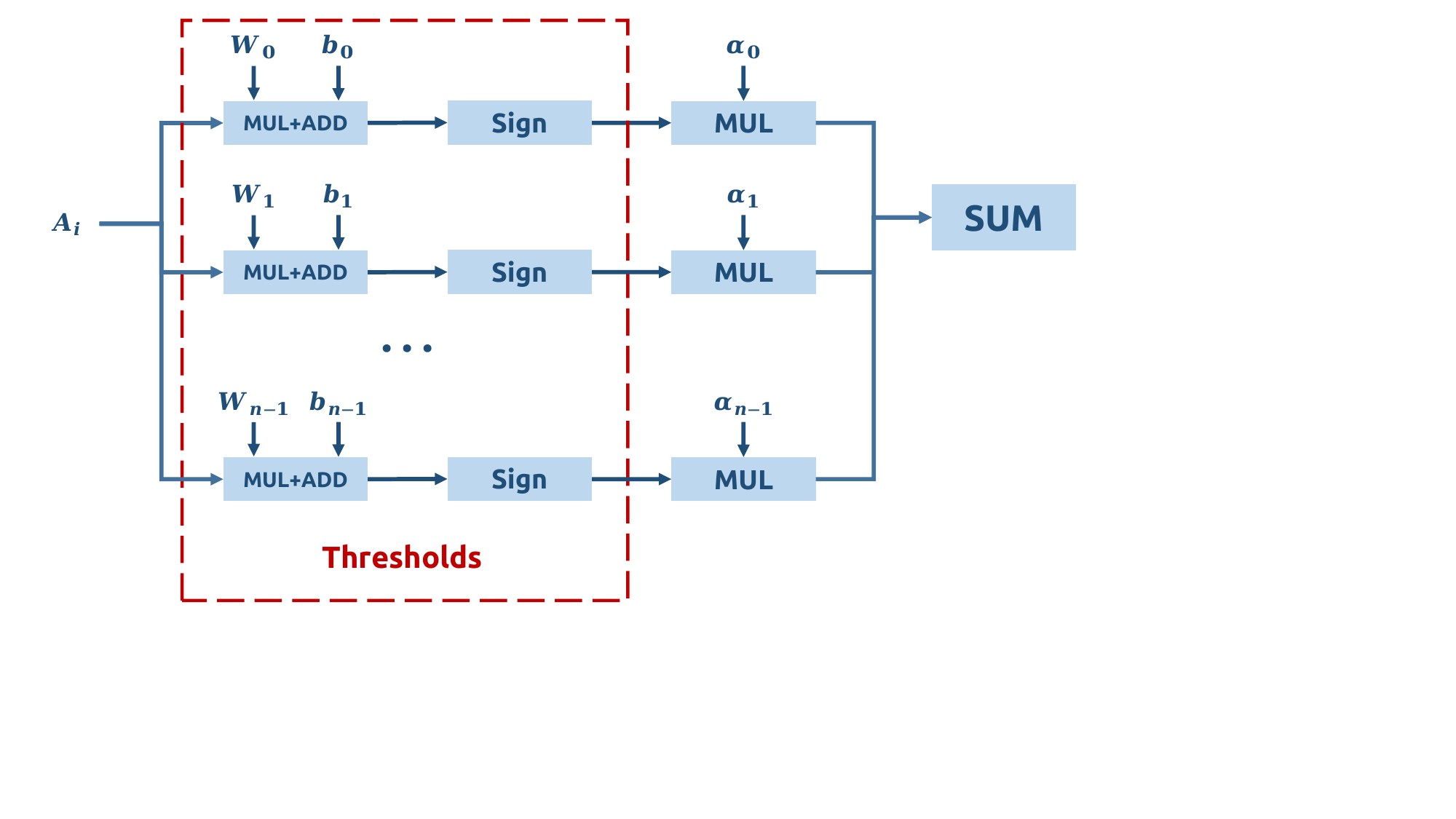}}
    \end{minipage}
    \caption{Converting one learnable nonlinear function to a series of weighted learnable thresholds}\
    \vspace{0pt}
    \label{WeightedThres}
    \begin{minipage}{1\columnwidth}
        \centerline{\includegraphics[width=0.9\columnwidth]{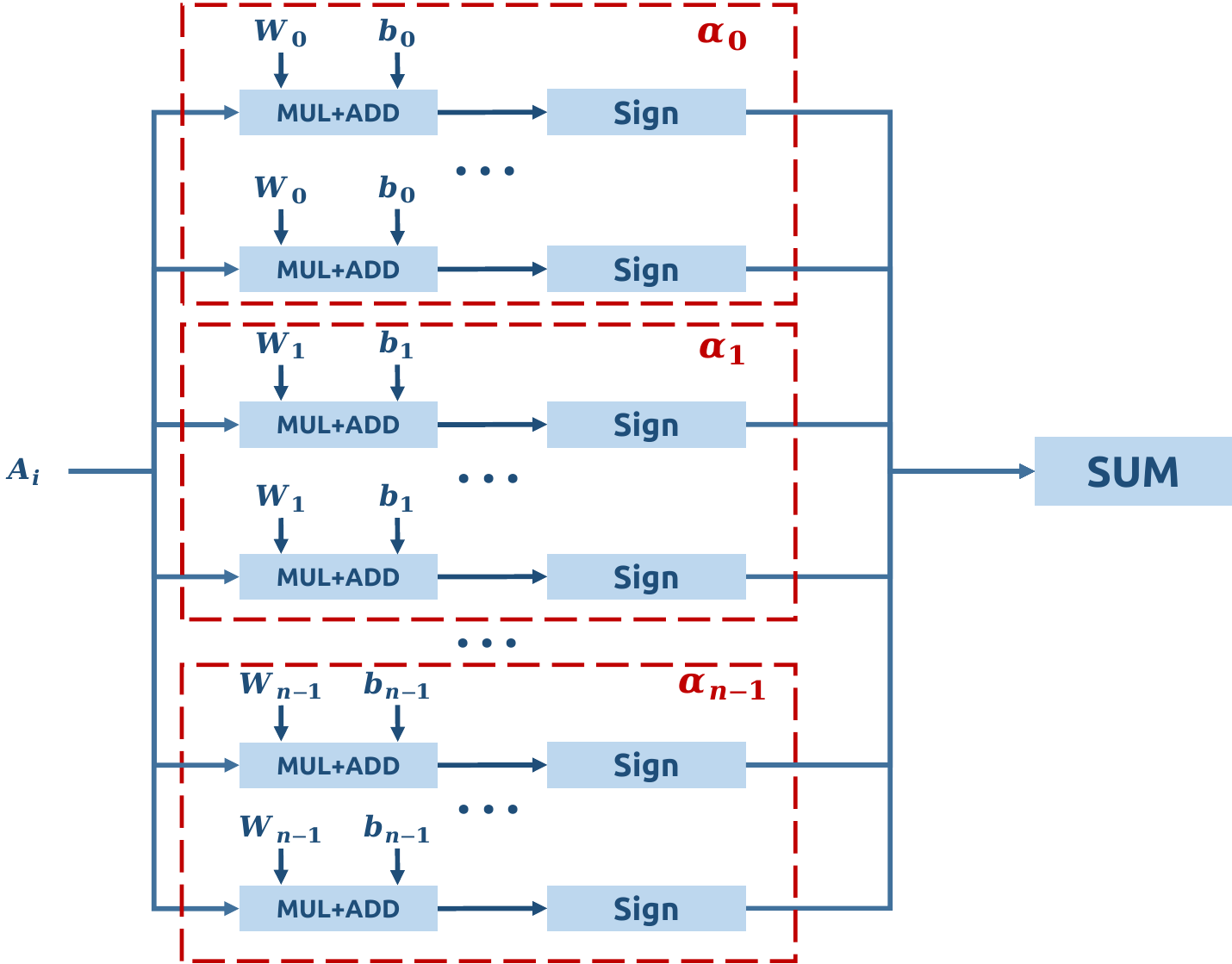}}
    \end{minipage}
    \caption{Duplicated learnable thresholds when all $\alpha_i$ are approximated as integers}
    \label{ApproxWeightedThres}
\end{figure}

As shown in~\autoref{Oiandweights}, for our constructed function $f'(x)$, when input is $s_i\leqslant x< s_{i+1}$, we can compute the result of each weighted threshold function. Therefore, for the weighted threshold functions, $\alpha_0Thres_{0}(x)$ to $\alpha_iThres_{i}(x)$, because their threshold values are $s_0$ to $s_i$ and the input is $x \geqslant s_i$, based on~\autoref{WeightedThres}, the output for these weighted thresholds are $\alpha_0$ to $\alpha_i$. According to the same principle, for the weighted threshold functions, $\alpha_{i+1}Thres_{i+1}(x)$ to $\alpha_{t-1}Thres_{t-1}(x)$, because their threshold values are $s_{i+1}$ to $s_{t-1}$ and the input is $x<s_{i+1}$, the output for these weighted thresholds are $-\alpha_{i+1}$ to $-\alpha_{t-1}$. Therefore, the sum of all weighted thresholds is $f'(x)=\sum_{i=0}^{t-1}\alpha_iThres_{i}(x)=\sum_{l=0}^{i}\alpha_l-\sum_{r=i+1}^{t-1}\alpha_r$. Because as shown in~\autoref{piecewise_constant}, when $s_i\leqslant x< s_{i+1}$, $f(x)=O_i$, we assume here the $O_i$ can be approximated as $O_i=\sum_{l=0}^{i}\alpha_l-\sum_{r=i+1}^{t-1}\alpha_r$.

\begin{equation}
    \small
    \centering
    \begin{aligned}
        & \begin{cases}
            O_0 = \alpha_0 - \sum_{j=1}^{t-1}\alpha_j & s_0\leqslant x<s_{1}\\
            O_1 = \alpha_0 + \alpha_1 - \sum_{j=2}^{t-1}\alpha_j & s_1\leqslant x<s_{2}\\
            O_2 = \alpha_0 + \alpha_1 + \alpha_2 - \sum_{j=3}^{t-1}\alpha_j & s_2\leqslant x<s_{3}\\
            \cdots \\
            O_{t-2} = \sum_{j=0}^{t-2}\alpha_j - \alpha_{t-1} & s_{t-2}\leqslant x<s_{t-1}\\
            O_{t-1} = \sum_{j=0}^{t-1}\alpha_j & s_{t-1}\leqslant x <s_{t}\\
          \end{cases} 
    \end{aligned}
    \label{BiKAFormula_1}
\end{equation}

Therefore, as shown in~\autoref{BiKAFormula_1}, we traverse input $x$ with all input slot $s_i\leqslant x<  s_{i+1}$ $(0\leqslant i\leqslant t-1)$ to compute the conversion between $O_i$ and $\alpha_i$ based on the formula of $O_i=\sum_{l=0}^{i}\alpha_l-\sum_{r=i+1}^{t-1}\alpha_r $.

\begin{equation}
    \small
    \centering
    \begin{aligned}
        & \begin{aligned}
            \begin{cases}
                O_0 = \alpha_0 - \sum_{j=0}^{t-1}\alpha_j + \alpha_0\\
                O_1 = \alpha_0 + \alpha_1 - \sum_{j=1}^{t-1}\alpha_j + \alpha_1\\
                \cdots \\
                \begin{aligned}
                O_{t-2} & = \sum_{j=0}^{t-3}\alpha_j + \alpha_{t-2} - \alpha_{t-2} - \alpha_{t-1} + \alpha_{t-2}
                \end{aligned} \\
                \begin{aligned}
                O_{t-1} & = \sum_{j=0}^{t-2}\alpha_j + \alpha_{t-1} - \alpha_{t-1} + \alpha_{t-1}
                \end{aligned}
            \end{cases}
        \end{aligned} \\
        \Rightarrow & \begin{aligned}
            \begin{cases}
                O_0 = 2\alpha_0 - O_{t-1} \\
                O_1 = 2\alpha_1 + O_{0}\\
                \cdots \\
                \begin{aligned}
                O_{t-2} & = 2\alpha_{t-2} + O_{t-3}
                \end{aligned} \\
                \begin{aligned}
                O_{t-1} & = 2\alpha_{t-1} + O_{t-2}
                \end{aligned}
            \end{cases}
        \end{aligned} \\
        \Rightarrow & \begin{aligned}
            \begin{cases}
                \alpha_0 = (O_0+O_{t-1})/{2} \\ 
                \alpha_1 = (O_1-O_0)/2 \\
                \cdots \\
                \alpha_{t-2} = (O_{t-2}-O_{t-3})/{2} \\ 
                \alpha_{t-1} = (O_{t-1}-O_{t-2})/{2} \\
            \end{cases}
        \end{aligned}
    \end{aligned}
    \label{BiKAFormula_2}
\end{equation}

Then, by substituting the expressions of 
$O_{t-1}$ in~\autoref{BiKAFormula_1} down to $O_0$ in~\autoref{BiKAFormula_2}, and $O_{i}$ in~\autoref{BiKAFormula_1} into $O_{i+1}$ in~\autoref{BiKAFormula_2} for $0 \leqslant i \leqslant t-2$, we obtain 
a closed-form solution for each threshold weight:
\begin{equation}
\small
\alpha_0 = \frac{O_0 + O_{t-1}}{2}, \qquad 
\alpha_i = \frac{O_i - O_{i-1}}{2}, \; 1 \leqslant i \leqslant t-1.
\label{finalweights}
\end{equation}

Since all $O_i$ values are fixed constants determined by the original function 
$f(x)$, we can compute every $\alpha_i$ directly. Therefore, the discrete nonlinear 
function $f(x)$ can be approximated by the weighted-threshold-based form as shown in~\autoref{constructedF}.

\begin{figure}[t]
    \centering
    \begin{minipage}{1\columnwidth}
        \centerline{\includegraphics[width=\columnwidth]{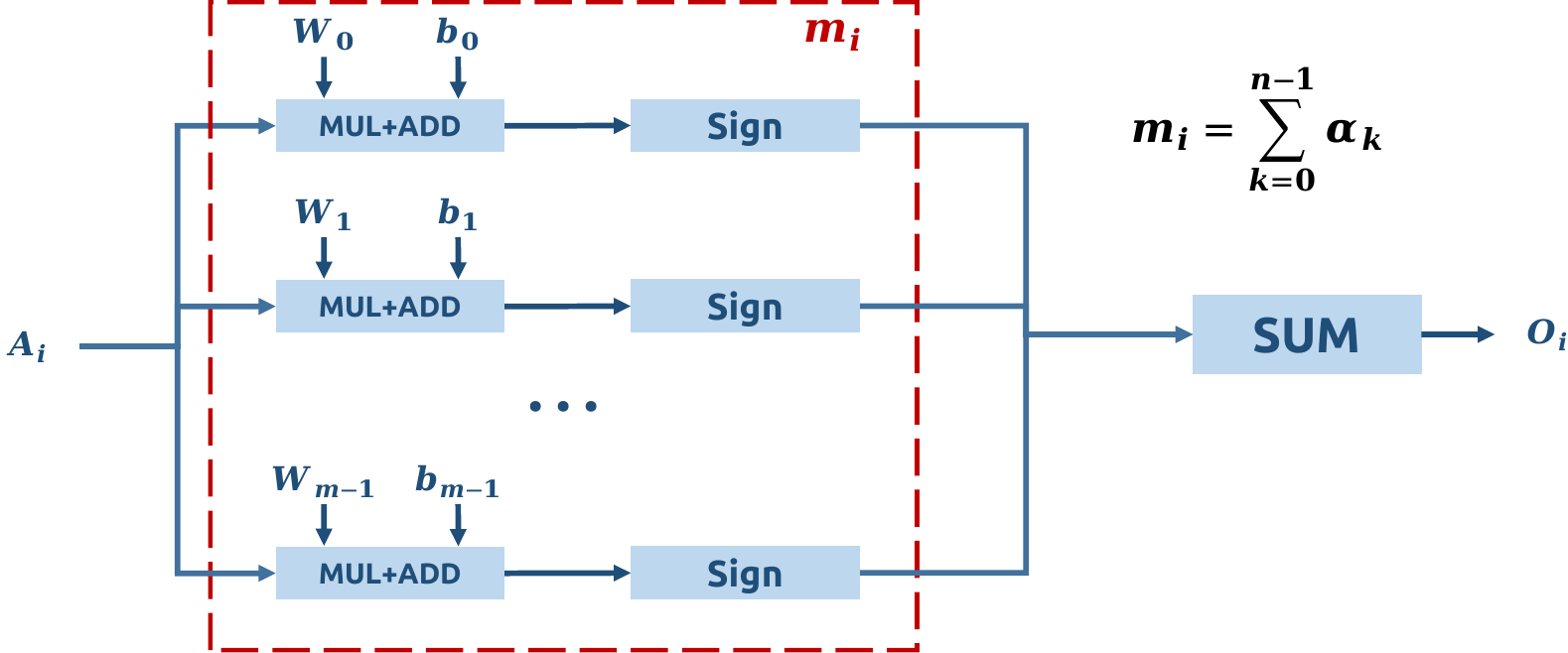}}
    \end{minipage}
    \caption{Mixed all learnable thresholds with the sum of $\alpha$}\
    \vspace{0pt}
    \label{MixedThres}
    \begin{minipage}{1\columnwidth}
        \centerline{\includegraphics[width=\columnwidth]{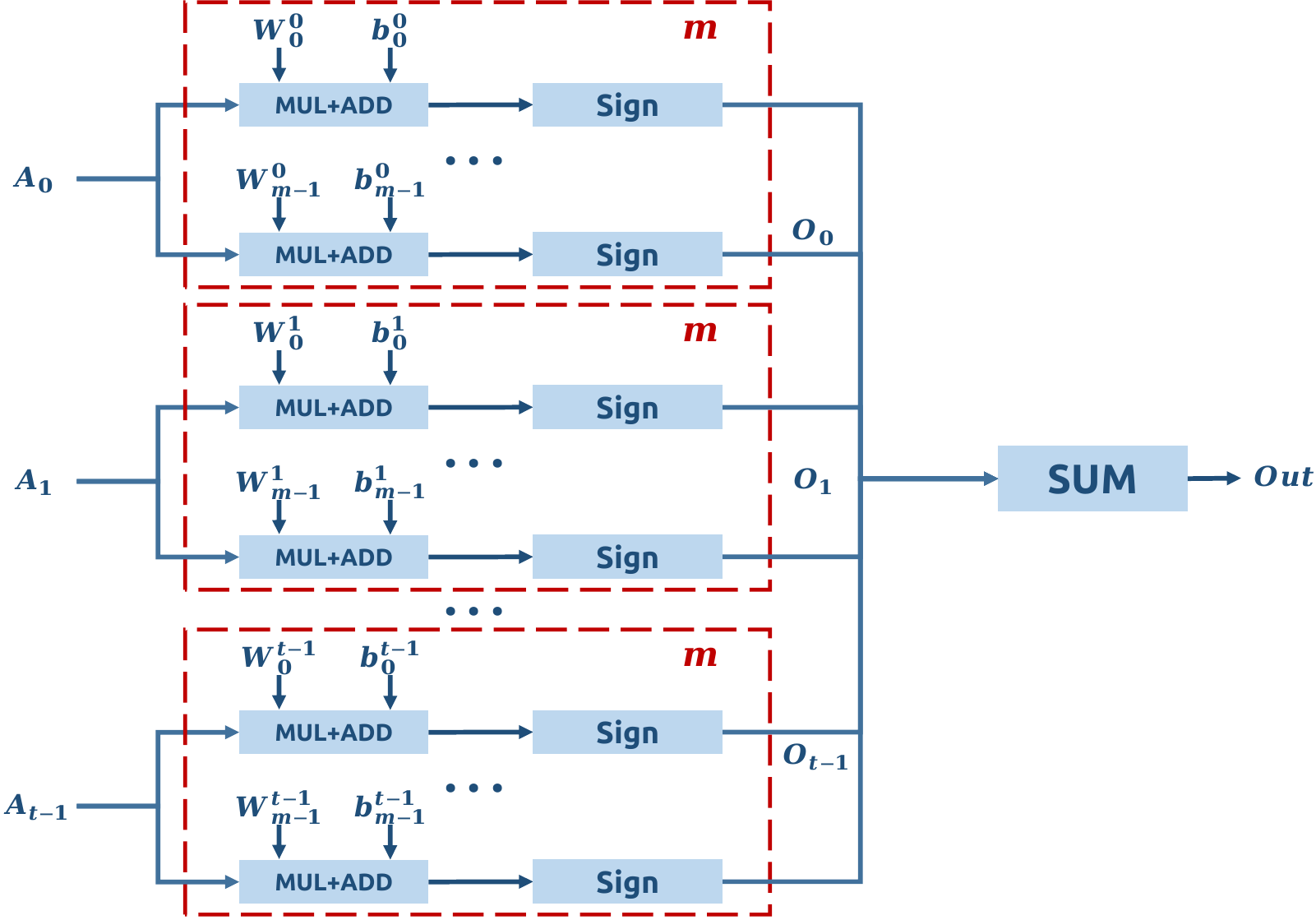}}
    \end{minipage}
    \caption{Setting $m$ as a quantized parameter to approximate and simplify learnable thresholds}\
    \label{UnifiedMixedThres}
\end{figure}

\subsection{Simplification of Approximate Nonlinear Functions for BiKA}

Considering the implementation of BNN in the FINN framework~\cite{Umuroglu2017, Blott2018}, we can convert a learnable threshold $Thres_i(x)$ as a combination of weight $w$, bias $\beta$, and \emph{Sign} activation as shown in~\autoref{learnablethres}. $-\frac{\beta}{w}$ is learned threshold value.

\begin{equation}
    \small
    \begin{aligned}
        Sign(w x+\beta ) & =
        \left\{
        \begin{matrix}
        1 & w x+\beta\geqslant 0\\
        -1 & w x+\beta < 0
        \end{matrix}\right. \\
        \Rightarrow 
        Sign(x) & =
        \left\{
        \begin{matrix}
        1 & x\geqslant -\frac{\beta}{w}\\
        -1 & x< -\frac{\beta}{w}
        \end{matrix}\right.
    \end{aligned}
    \label{learnablethres}
\end{equation}

Therefore, based on the above-discussed mathematics conversion, ~\autoref{WeightedThres} shows that we can convert the training of one learnable nonlinear function $f(x)$ as a series of parameters of weights $w$, biases $\beta$, \emph{Sign} activations, and threshold weights $\alpha$ with two multipliers and one adder.

However, this conversion of one nonlinear function remains too complex and hardware-expensive due to the dense number of parameters and computation. To simplify it, if we approximate the $O_0+O_{t-1}$ and $O_i-O_{i-1}$ in~\autoref{finalweights} as even integers, the $\alpha_i$ values of $\frac{O_0+O_{n-1}}{2}$ and $\frac{O_i-O_{i-1}}{2}$ are integers too. Therefore, as shown in~\autoref{ApproxWeightedThres}, multiplying the output of the \emph{Sign} activation by threshold weights $\alpha_i$ can be simplified by duplicating $\alpha_i$ times of the input to remove one multiplier for the threshold weight. As a result, comparing \autoref{WeightedThres} and \autoref{ApproxWeightedThres}, one weighted learnable threshold is converted into $\alpha_i$ times of learnable thresholds.

Considering the order of thresholds will not influence the final output, we can define an integer number $m$ as the sum of all quantized $\alpha_i$ to mix all thresholds converted from different weighted thresholds in one nonlinear function together, as shown in~\autoref{MixedThres}. Therefore, a higher $m$ can represent the nonlinear function more accurately. Then, for further approximation, we set $m$ as a unified parameter for all nonlinear functions in our approximated KAN as shown in~\autoref{MixedThres}. As a result, the output range of each $O_i$ in~\autoref{MixedThres} is $[-m,m]$. Therefore, the inputs, outputs, and thresholds of every layer and neuron in this approximated KAN model are quantized as integers, adjusted by the quantization factor $m$. 

\begin{figure}[t]
    \centering
    \begin{minipage}{1\columnwidth}
        \centerline{\includegraphics[width=\columnwidth]{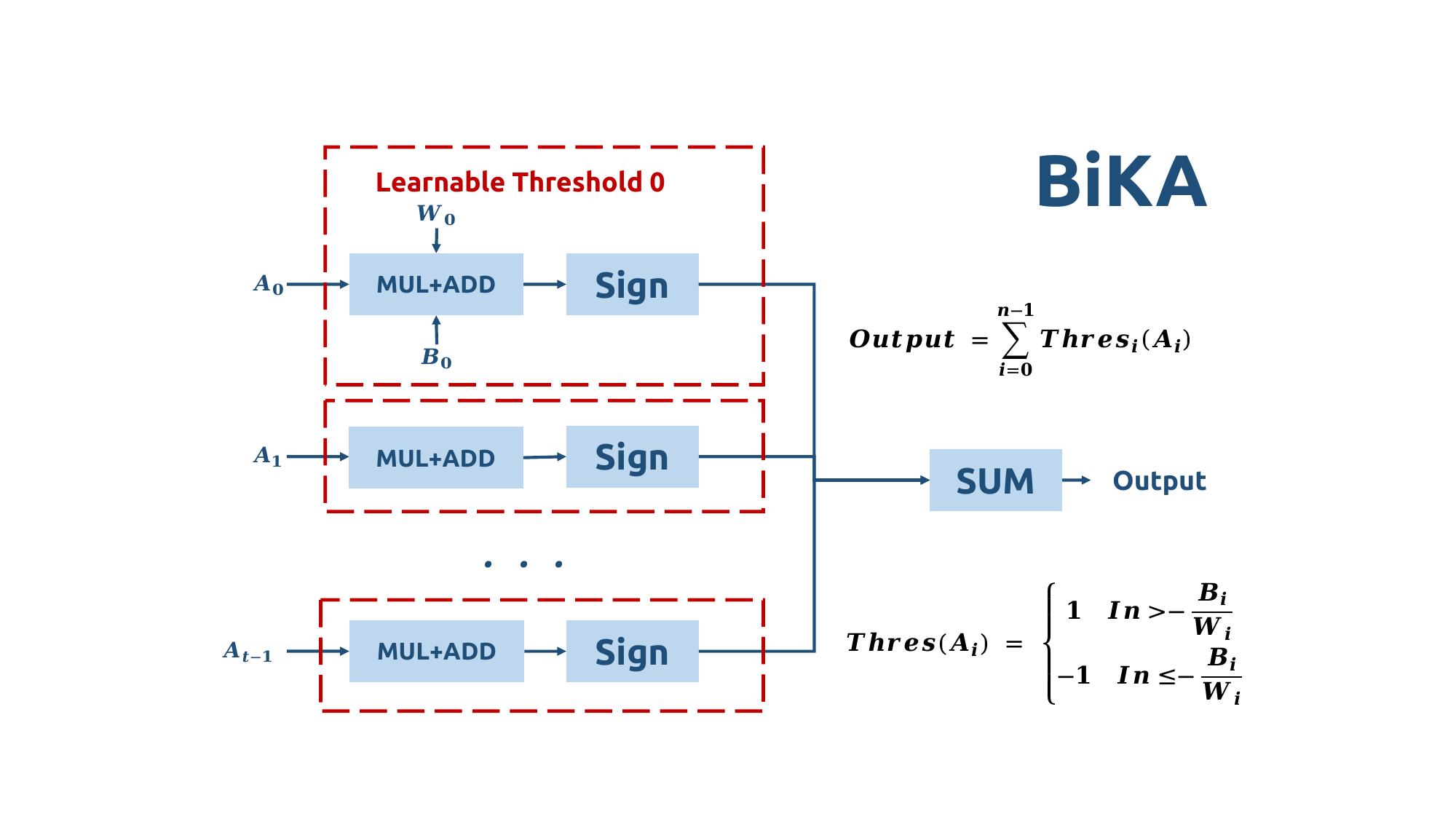}}
    \end{minipage}
    \caption{Computation in BiKA neurons when set quantization parameter, $m$, as one.}\
    \label{BinKAN}
    \begin{minipage}{1\columnwidth}
        \centerline{\includegraphics[width=\columnwidth]{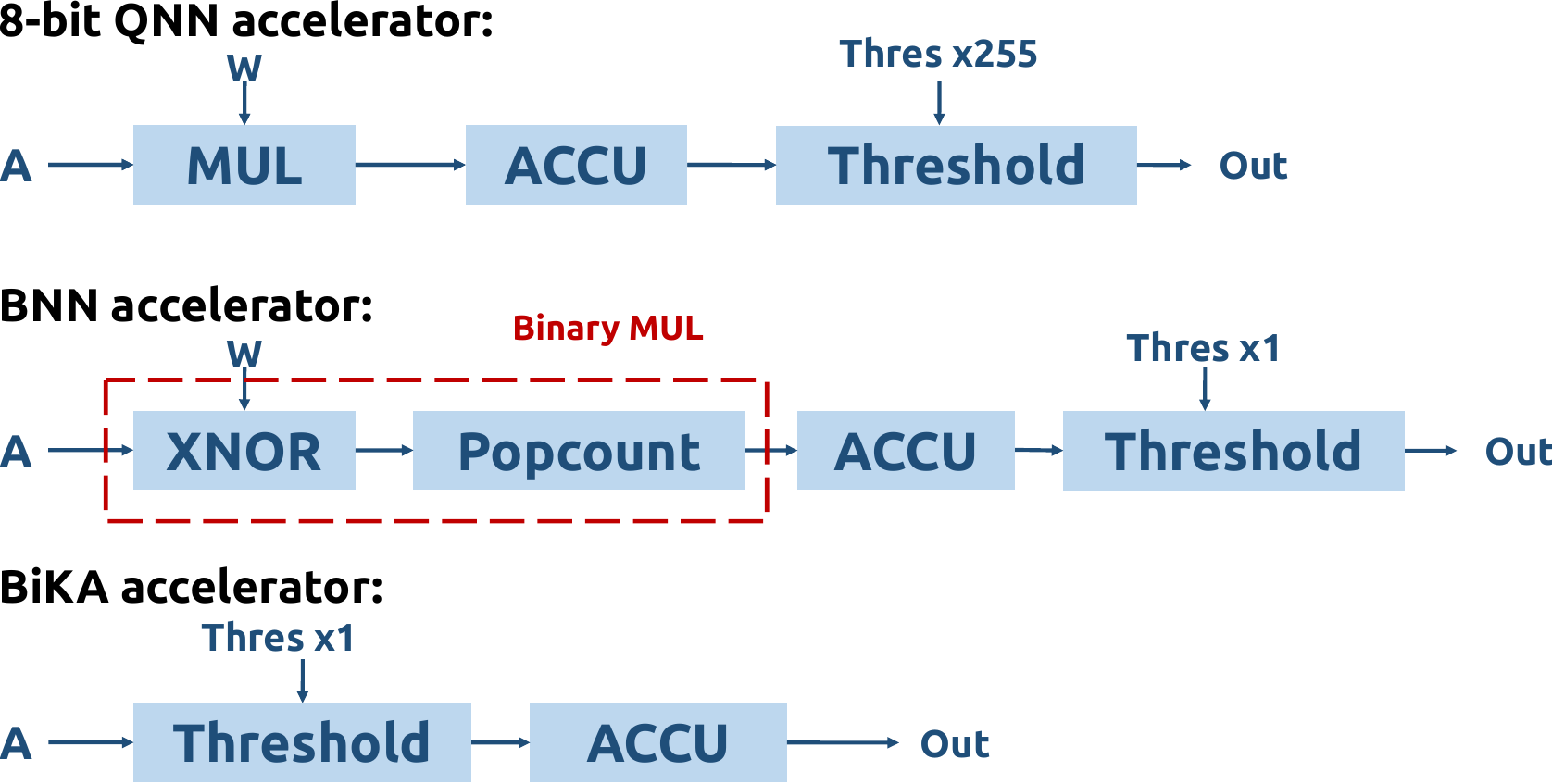}}
    \end{minipage}
    \caption{Hardware difference in accelerator designs between BNN, QNN, and BiKA networks}\
    \label{NeuronHardware}
\end{figure}

This approximation method has an extreme situation: when $m$ is set to one, each learnable nonlinear activation function will be approximated as a single learnable threshold. We named this situation \emph{Binarized KAN} (BiKA) as shown in~\autoref{BinKAN}. 

In the traditional ANN models, the computation in one neuron or kernel is $ReLU(\sum_{i=0}^{n-1} W_i a_i+b)$, where $W_i$ is the weight, $a_i$ is the input activation, and $b$ is the bias. In our BiKA, the computation in one neuron or kernel is $\sum_{i=0}^{n-1} Sign(W_i a_i+b_i)$, where $W_i$ is the weight, $a_i$ is the input activation, and $b_i$ is the individual bias for each input activation. $Sign(W_i a_i+b_i)$ can be converted into one threshold. Therefore, this architecture can be customized for training based on \emph{PyTorch} and CUDA. We implemented a customized training library for BiKA by introducing \emph{BiKALinear} and \emph{BiKAConv2d}, which modify the standard \emph{Linear} and \emph{Conv2d} layers to match BiKA’s threshold-based computation pattern. Because BiKA relies heavily on Sign threshold operations, the true gradient is not available. To enable stable backpropagation, we replace the backward pass of the Sign function with the derivative of the hard-tanh function in our CUDA implementation. This straight-through estimator greatly improves the training stability and accuracy of the BiKA models.

\subsection{Processing Element and Systolic Array Design of BiKA}

\begin{figure}[t]
    \centering
    \begin{minipage}{1\columnwidth}
        \centerline{\includegraphics[width=\columnwidth]{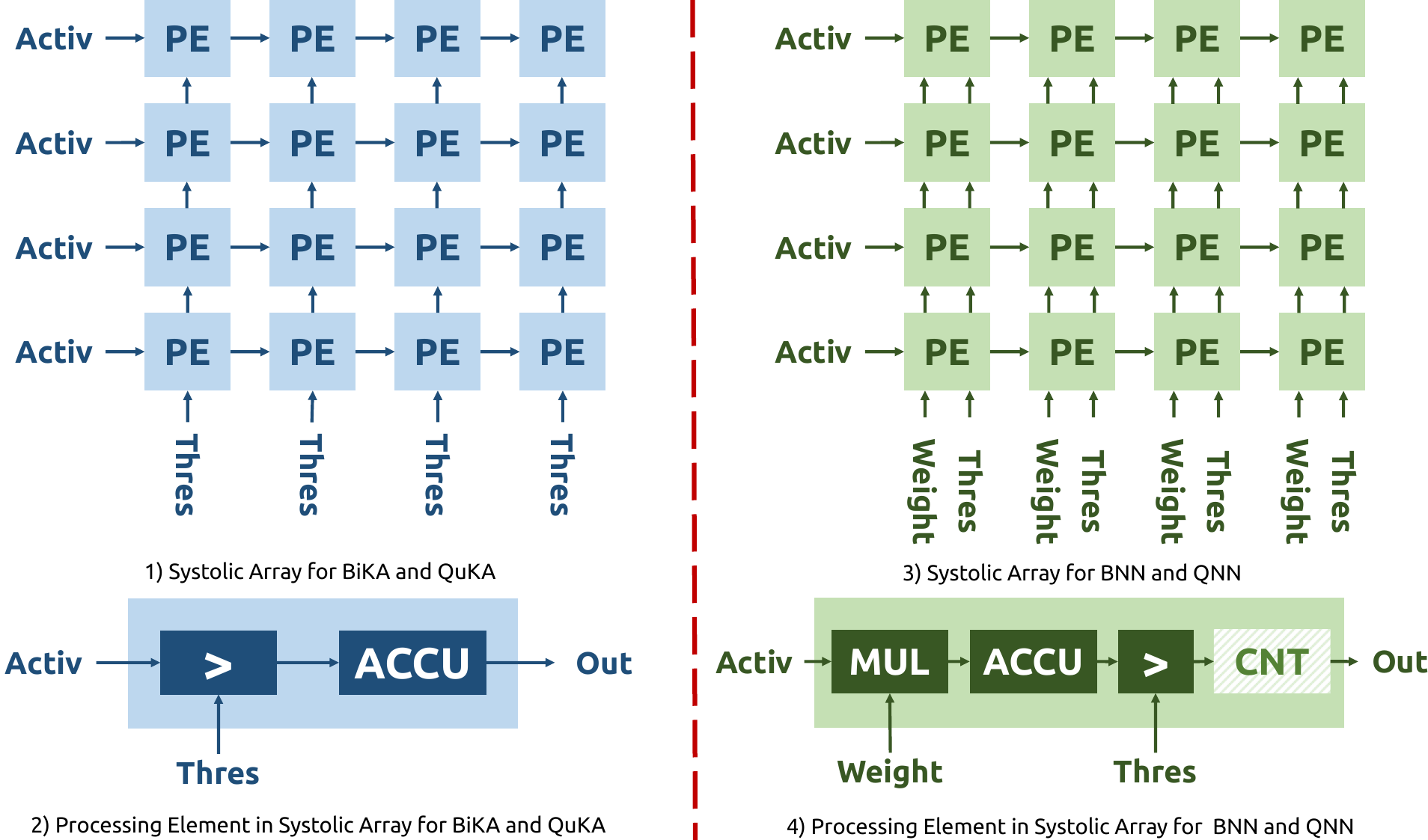}}
    \end{minipage}
    \caption{Hardware difference of systolic array accelerators between BNN, QNN, and BiKA}\
    \label{SystolicHardware}
\end{figure}

Considering the above-discussed designs of BiKA networks, we present and compare the different hardware designs between QNN, BNN, and BiKA in this manuscript. We implemented three systolic array accelerators for QNN, BNN, and BiKA. Each systolic array consists of $8\times8$ processing elements as shown in \autoref{SystolicHardware}. Both BNN and QNN accelerators implemented in this work are inspired by~\emph{FINN}~\cite{Umuroglu2017, Blott2018}. The original FINN framework is designed for the data-streaming architecture of FPGA-based NN accelerators. We expand its design as systolic arrays. 

Compared to BNN, QNN, and our BiKA processing elements shown in \autoref{NeuronHardware}, the major differences are: 
\begin{itemize}
    \item Both BNN and QNN processing elements need threshold-based activation modules. The BNN processing element only has one threshold. The $n$-bit QNN processing element requires $2^n$ thresholds for activation and quantization in output according to \emph{FINN-R}~\cite{Blott2018}. To reduce the hardware resource consumption, we implement only one comparator in the QNN accelerator and process all threshold comparisons in serial.
    \item Because BNN represents $-1$ as 1-bit '0' and $1$ as 1-bit '1', the computation in BNN is XNOR. Moreover, because the BNN accelerator computes multi-bits in parallel to speed up the inference, one \emph{PopCount} module is used to compute the sum of XNOR output inspired by~\emph{FINN}~\cite{Umuroglu2017}.
    \item Compared with the processing element design of BNN and QNN, the BiKA processing element only needs to implement one comparator to replace the multiplier and XNOR+\emph{PopCount} unit without the additional threshold activations as in BNN and QNN, which can highly simplify the design of the systolic array engine and the control state machine in the BiKA accelerators.
\end{itemize}
 
\begin{table*}
    \centering
    \caption{Inference Accuracy of BNN, QNN, KAN, and BiKA on MLP and tiny CNN with MNIST Classification}
    \resizebox{2\columnwidth}{!}
    {
        \begin{tabular}{ccccc}
            \toprule
            Dataset                  & \multicolumn{3}{c}{MNIST}~\cite{deng2012mnist}                                                                                   & CIFAR-10~\cite{Krizhevsky2009}                              \\ \midrule
            \multirow{3}{*}{Network} & TFC                          & SFC                                 & LFC                                    & CNV                                   \\
                                     & \multirow{2}{*}{F64/F32/F10} & \multirow{2}{*}{F256/F256/F256/F10} & \multirow{2}{*}{F1024/F1024/F1024/F10} & C64/C64/P2/C128/C128/P2/C256/C256/P2/ \\ 
                                     &                              &                                     &                                        & F512/F512/F10                         \\ \midrule
            Q-KAN~\cite{qkan}                    & 98.52\%                      & -                                   & -                                      & -                                     \\
            KAN~\cite{qkan}                      & 98.02\%                      & -                                   & -                                      & -                                     \\ \midrule
            BNN                      & 93.34\%                      & 97.39\%                             & 98.05\%                                & 65.20\%                               \\
            QNN                      & 97.92\%                      & 98.59\%                             & 98.68\%                                & 72.76\%                               \\
            KAN                      & 96.14\%                      & 95.83\%                             & -                                      & -                                     \\
            BiKA                     & 91.94\%                      & 96.46\%                             & 97.86\%                                & 55.80\%                              \\ \bottomrule
        \end{tabular}
    }
    \label{train_tab}
\end{table*}

\begin{figure*}[t]
\centering
{
    \begin{minipage}{1\columnwidth}
        \centering
        \begin{tikzpicture}
            \begin{axis}[
                width=8cm,
                height=8cm,
                scale=0.95,
                view={0}{90},
                axis equal image,       
                xmin=0.5, xmax=8.5,    
                ymin=0.5, ymax=3.5,      
                enlargelimits=false,
                axis on top,
                axis lines=box,
                xtick={1, 2, 3, 4, 5, 6, 7, 8},
                ytick={1, 2, 3},
                xticklabels={
                    {A},
                    {B},
                    {C},
                    {D},
                    {E},
                    {F},
                    {G},
                    {H}
                },
                yticklabels={
                    {Batch 1024},
                    {Batch 512},
                    {Batch 256}
                },
                ticklabel style={font=\scriptsize},
                colormap/viridis,
                colorbar,
            ]
            
            \addplot[
                matrix plot*,
                mesh/cols=8,
                point meta=explicit,
                point meta min=92,
                point meta max=97.77,
                draw=white,
                nodes near coords={%
                    \contour{white}{\pgfmathprintnumber[fixed,precision=2]{\pgfplotspointmeta}}%
                },
                every node near coord/.style={
                    font=\scriptsize,
                    text=black  
                },
            ]
            table[meta=z]{
            x y z
            1 1 96.05
            2 1 97.49
            3 1 97.86
            4 1 97.57
            5 1 97.11
            6 1 97.77
            7 1 97.23
            8 1 97.16
            1 2 91.55
            2 2 96.50
            3 2 96.66
            4 2 96.84
            5 2 97.13
            6 2 97.77
            7 2 97.21
            8 2 97.20
            1 3 80.32
            2 3 92.62
            3 3 92.34
            4 3 93.76
            5 3 95.03
            6 3 97.51
            7 3 97.44
            8 3 97.19
            };
            
            \end{axis}
        \end{tikzpicture} 
    \end{minipage}
    \begin{minipage}{1\columnwidth} 
        \centering
        \begin{tikzpicture}
            \begin{axis}[
                width=8cm,
                height=8cm,
                scale=0.95,
                view={0}{90},
                axis equal image,       
                xmin=0.5, xmax=8.5,    
                ymin=0.5, ymax=3.5,      
                enlargelimits=false,
                axis on top,
                axis lines=box,
                xtick={1, 2, 3, 4, 5, 6, 7, 8},
                ytick={1, 2, 3},
                xticklabels={
                    {A},
                    {B},
                    {C},
                    {D},
                    {E},
                    {F},
                    {G},
                    {H}
                },
                yticklabels={
                    {Batch 1024},
                    {Batch 512},
                    {Batch 256}
                },
                ticklabel style={font=\scriptsize},
                colormap/viridis,
                colorbar,
            ]
            
            \addplot[
                matrix plot*,
                mesh/cols=8,
                point meta=explicit,
                point meta min=30,
                point meta max=54,
                draw=white,
                nodes near coords={%
                    \contour{white}{\pgfmathprintnumber[fixed,precision=2]{\pgfplotspointmeta}}%
                },
                every node near coord/.style={
                    font=\scriptsize,
                    text=black
                },
            ]
            table[meta=z]{
            x y z
            1 1 45.63
            2 1 54.15
            3 1 54.21
            4 1 53.72
            5 1 48.44
            6 1 49.31
            7 1 52.63
            8 1 51.79
            1 2 35.64
            2 2 46.33
            3 2 45.29
            4 2 43.05
            5 2 47.38
            6 2 53.56
            7 2 53.43
            8 2 52.83
            1 3 30.25
            2 3 39.97
            3 3 39.97
            4 3 38.97
            5 3 47.93
            6 3 54.55
            7 3 55.8
            8 3 53.95
            };
            
            \end{axis}
        \end{tikzpicture} 
    \end{minipage}
}
\begin{minipage}{2\columnwidth}
    \centering
    \begin{tikzpicture}
        \node[draw, rounded corners, inner sep=6pt, align=left, scale=0.85, font=\scriptsize] (legend) {
        \textbf{Learning Rate Combinations:}\\[4pt]
        A: $LR_0$=0.0010, $LR_1$=0.0010, $LR_2$=0.0010 \hspace{5pt}
        B: $LR_0$=0.0010, $LR_1$=0.0005, $LR_2$=0.0002 \hspace{5pt} 
        C: $LR_0$=0.0010, $LR_1$=0.0005, $LR_2$=0.0001 \hspace{5pt} 
        D: $LR_0$=0.0010, $LR_1$=0.0002, $LR_2$=0.0001\\
        E: $LR_0$=0.0005, $LR_1$=0.0005, $LR_2$=0.0005 \hspace{5pt}
        F: $LR_0$=0.0005, $LR_1$=0.0002, $LR_2$=0.0001 \hspace{5pt}
        G: $LR_0$=0.0002, $LR_1$=0.0002, $LR_2$=0.0002 \hspace{5pt}
        H: $LR_0$=0.0001, $LR_1$=0.0001, $LR_2$=0.0001\\
        };
    \end{tikzpicture}
\end{minipage}

\caption{Accuracy in MNIST (left) and CIFAR-10 (right) datasets based on LFC and CNV models with different hyperparameters}
\label{heatingmap}
\end{figure*}
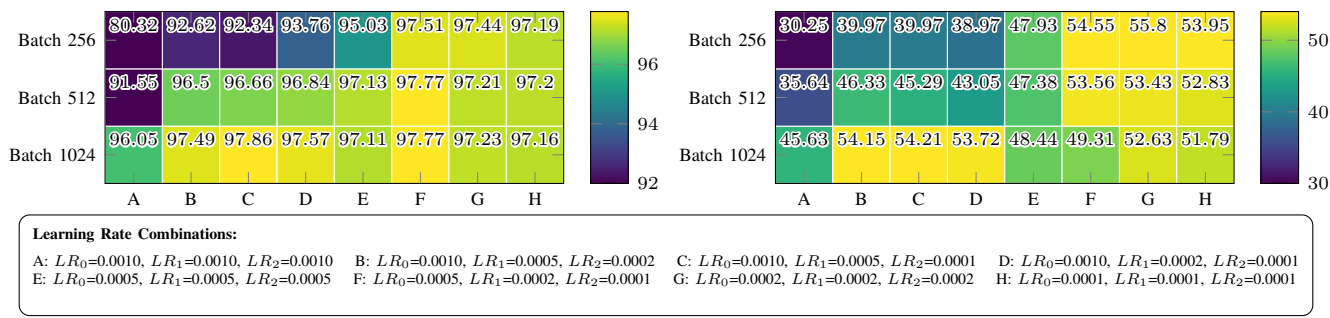

\begin{figure*}[t]
    \centering
    \begin{minipage}{1\columnwidth}
        \centerline{\includegraphics[width=\columnwidth]{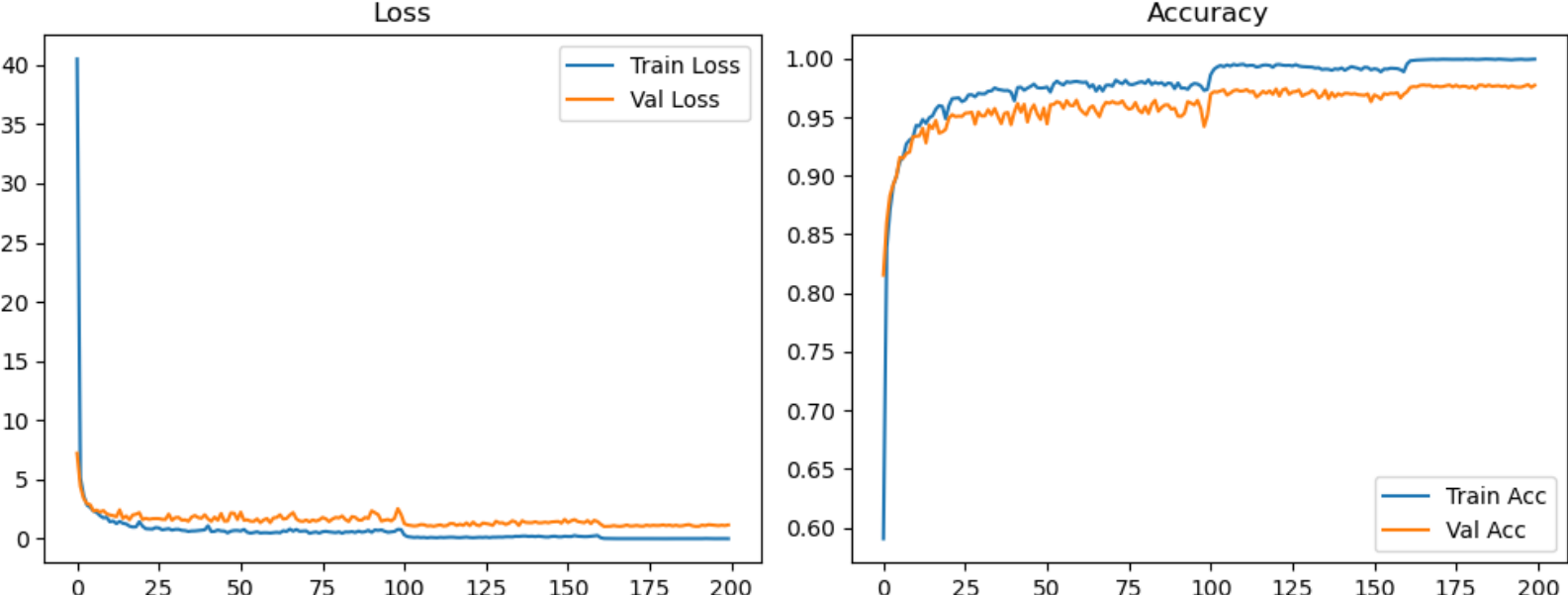}}
        \vspace{5pt}
    \end{minipage}
    \begin{minipage}{1\columnwidth}
        \centerline{\includegraphics[width=\columnwidth]{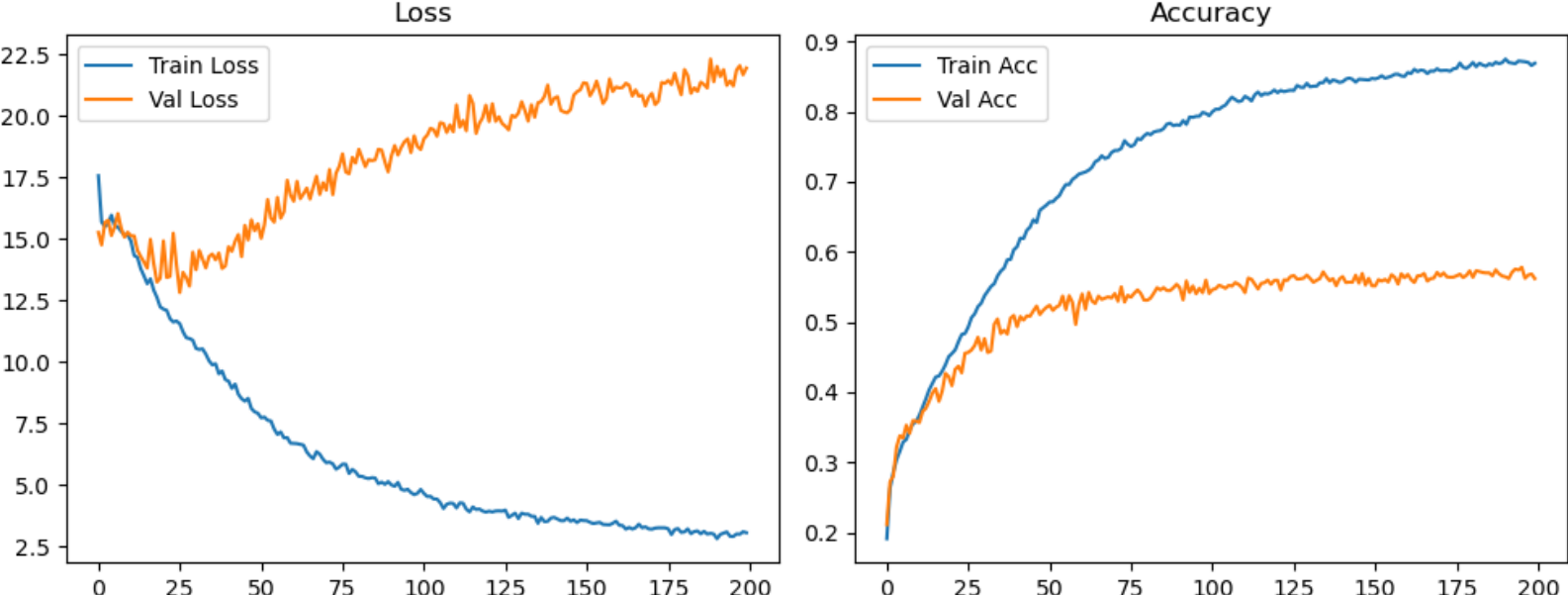}}
        \vspace{5pt}
    \end{minipage}
    \caption{Accuracy and loss in training and validation based on LFC and CNV models with MNIST (left) and CIFAR-10 (right)}
    \label{training_plot}
\end{figure*}

Based on the processing element shown in~\autoref{NeuronHardware}, we implement systolic-array-based approaches in this work to achieve higher scalability for larger network models. Compared to BNN/QNN and our BiKA systolic array engine designs shown in \autoref{SystolicHardware}, the major differences of our BiKA accelerator are: 

\begin{itemize}
    \item In our BiKA accelerator, thresholds replaced the weights to execute the systolic array processing. Therefore, \emph{Multiply-Accumulators} (MACs) have been replaced with \emph{Comparison-Accumulators} (CACs).
    \item In our BNN/QNN accelerator, we implement one additional pipeline in the systolic array to reuse the loaded threshold for processing elements and improve the throughput. After finishing the multiply-accumulation computing in all processing elements, our BNN/QNN systolic array will idle the MACs and load the thresholds for nonlinear activation and output quantization processing. This implementation is unnecessary in our BiKA accelerator, since BiKA doesn't need the additional nonlinear activation after CAC computation.
\end{itemize}

%% file: ISQED2026/evaluation.tex
\subsection{Training Experiment of BiKA}\begin{table*}
    \centering
    \caption{Resource Consumption of BNN, QNN, and BiKA Accelerators on \textit{Ultra96V2} FPGA Platform}
    \resizebox{2\columnwidth}{!}
    {
        \begin{tabular}{cccccccccccc}
            \toprule
            \multirow{2}{*}{Accelerator} & \multicolumn{3}{c}{Hardware Resource} & Frequency & Total Delay & ADP       & Dynamic Power & PDP    & \multicolumn{3}{c}{Latency ($\mu s$)}               \\
                                         & LUT         & FF          & BRAM      & (MHz)     & (ns)        & (ns)      & (W)           & (W*ns) & TFC          & SFC          & LFC         \\ \midrule
            QKAN-max~\cite{qkan}                     & $>$65680      & -           & -         & -         & -           & -         & -             & -      & -            & -            & -           \\
            QKAN-min~\cite{qkan}                     & $>$28371      & -           & -         & 100       & -           & -         & 0.612         & -      & 4.74         & -            & -           \\ \midrule
            BiKA                         & 8900        & 9232        & 19.5      & 300       & 2.744       & 24421.600 & 1.778         & 4.878  & 11.201    & 71.421        & 611.890            \\
            BNN                          & 12315       & 9962        & 24.5      & 300       & 3.013       & 37105.095 & 1.860         & 5.604  & 1.646     & 10.663     &    84.753         \\
            QNN                          & 18366       & 13179       & 23.5      & 250       & 3.610       & 66301.260 & 1.803         & 6,509  & 34.915    & 236.028   & 1327.980            \\ \bottomrule
        \end{tabular}
    }
    \label{hardware_tab}
\end{table*}

As shown in~\autoref{train_tab}, we trained 14 different models of BiKA, KAN, BNN, and 8-bit QNN networks for comparison based on four network structures: 

\begin{itemize}
    \item \emph{Tiny Fully Connected Network} (TFC), \emph{Small Fully Connected Network} (SFC), and \emph{Large Fully Connected Network} (LFC) are three MLP structures. TFC is a three-layer MLP structure that consists of 64/32/10 neurons, the same as the evaluated MLP network in work~\cite{qkan}. SFC and LFC are two four-layer MLP structures that consist of 256/256/256/10 neurons and 1024/1024/1024/10 neurons. All of these three MLP structures are used to evaluate the MNIST dataset~\cite{deng2012mnist} with 200 epochs of training.
    \item \emph{Convolution Neural Network} (CNV) is one tiny VGG-like CNN structure that consists of six convolutional layers and three fully connected layers. Two convolutional layers and one \emph{MaxPool} layer are combined as a block. This structure is used to evaluate the CIFAR-10~\cite{Krizhevsky2009} dataset with 200 epochs of training.
\end{itemize}

As shown in the second row of~\autoref{train_tab}, F64 means this layer is a fully connected layer with 64 neurons, and C64 and P2 are one convolutional layer with 64 $3\times3$ kernels with the padding of 1 and the stride of 1, and one 2x2 \emph{MaxPool} layer with the padding of 1 and the stride of 1.  

The third and fourth rows of \autoref{train_tab} list the reported training results of KAN and quantized KAN from Yin et al.~\cite{qkan}, showing $98.02\%$ for KAN and $98.52\%$ for their LUT-based quantized KAN on the MNIST~\cite{deng2012mnist} dataset using the TFC structure. As KAN training is known to be sensitive to initialization and hyperparameter choices, we additionally reproduced the TFC KAN model in our own environment for a consistent comparison against BNN, QNN, and BiKA. Since Yin et al.’s quantized KAN is directly derived from their trained KAN model, we expect its accuracy in our environment to be close to that of our reproduced KAN.

Therefore, the training results listed in~\autoref{train_tab} show that:

\begin{itemize}
    \item Due to the high memory usage of the native \emph{pykan} library during training, we only trained the TFC and SFC models for KAN. In our experiments, KAN achieves higher accuracy than BNN in the TFC structure, but becomes lower than BNN in the larger SFC structure.

    \item In the TFC model, KAN and 8-bit QNN outperform BNN by $2.8\%$ and $4.58\%$, respectively, while BiKA shows the lowest accuracy, $1.4\%$ below BNN. However, as the network size increases, BiKA starts to outperform KAN from the SFC model onward. The accuracy gaps between BNN and BiKA on MNIST also shrink to $0.93\%$ and $0.19\%$ in the SFC and LFC models. Since the quantized KAN of Yin et al.~\cite{qkan} is derived from their trained KAN model, we expect its accuracy in our environment to be close to that of our reproduced KAN. Under this assumption, BiKA achieves competitive accuracy compared with both native and quantized KAN.

    \item To evaluate BiKA on a more complex dataset, we additionally tested CIFAR-10. The last column of \autoref{train_tab} compares BNN, QNN, and BiKA, where BiKA shows a $9.4\%$ accuracy drop compared with BNN. To investigate this gap, we trained BiKA under various initial hyperparameter settings. As shown in \autoref{heatingmap}, BiKA is highly sensitive to the choice of batch size and learning rate, with accuracy changes up to $17.45\%$ on MNIST and $25.55\%$ on CIFAR-10~\cite{Krizhevsky2009}. The heat map also indicates that larger batch sizes and smaller learning rates generally yield better results. Therefore, a more refined training strategy is expected to improve BiKA’s CIFAR-10 accuracy.

    \item \autoref{training_plot} shows the training and validation curves for MNIST and CIFAR-10. BiKA achieves approximately $90\%$ training accuracy on CIFAR-10 but only about $55\%$ validation accuracy, while such divergence does not appear on MNIST. These observations suggest that BiKA has sufficient expressivity to fit the CIFAR-10 training data. However, the capacity of the lightweight CNV model is not enough for good generalization, leading to overfitting. Increasing the network size or applying regularization techniques, such as weight decay, is likely to further improve BiKA’s performance on CIFAR-10~\cite{Krizhevsky2009}.

\end{itemize}

\subsection{Hardware Accelerator Evaluation of BiKA}

As shown in~\autoref{hardware_tab}, we implemented four systolic array accelerators for BiKA, BNN, and QNN on \emph{Ultra96v2} FPGA platform with $8\times 8$ processing elements based on the design shown in~\autoref{SystolicHardware}. The QNN accelerator in the work is designed to infer the network models with 8-bit quantized activation and weights. The 8-bit BNN accelerator loads the 8-bit binarized inputs and weights in one shot to execute XNOR-Popcount computation eight times in parallel for higher throughput. For the BiKA accelerator, we implemented one 8-bit instance in this evaluation. As shown in~\autoref{BinKAN}, when outputs of the processing element are set as 8-bit, the output range of the accumulator is $[-128, 127]$, since the outputs of all threshold activations are $-1$ or $1$. Therefore, in principle, the 8-bit output of the accumulator can only support at most 127 inputs. However, in our experiments, we found that in most cases, the sum values in the accumulator of networks are not out of the range of $[-128, 127]$. Therefore, we implemented an 8-bit BiKA accelerator with a sum limitation design in the accumulator to prevent the value from being out of bounds. The hardware implementation results listed in~\autoref{hardware_tab} show that: 

\begin{itemize}
    \item \textbf{Hardware Resource Consumption}: The 8-bit BiKA accelerator significantly reduces LUT consumption compared to 8-bit BNN and QNN accelerators, by $27.73\%$ and $51.54\%$, respectively. This suggests our BiKA has the potential to implement the ultra-lightweight accelerator with scalability to support larger network inference based on the systolic-array structure. 
    \item \textbf{Clock Frequency}: All BiKA and BNN accelerators support a 300$MHz$ clock, higher than QNN accelerators. However, the total delay of BiKA accelerators reported from the implementation in \emph{Vivado} is lower than the BNN and QNN accelerators.
    \item \textbf{Aera-Delay-Products} and \textbf{Power-Delay-Products}: Furthermore, with the power, total delay, and LUT consumption reported from Vivado implementation, our BiKA has the lowest~\emph{Aera-Delay-Products} (ADP) and~\emph{Power-Delay-Products} (PDP) compared to BNN and QNN accelerators, which shows the design and power efficiency of BiKA accelerators.
    \item \textbf{Inference Latency}: We evaluated three network models, TFC, SFC, and LFC, on these four accelerators. Results show that our BiKA is $2.17\times$-$3.30\times$ faster than 8-bit QNN. However, the 8-bit \emph{Single Instruction/Multiple Data} (SIMD) design in the BNN systolic array engine has a significant speedup compared with other accelerators, which makes it the fastest accelerator implementation in our evaluation. However, the ultra-low hardware resource consumption of BiKA can be more suitable than the BNN accelerator in some extreme resource-limited, non-speed-sensitive scenarios. 
\end{itemize}

Moreover, we also compared our BiKA accelerator with the quantized KAN accelerator of Yin et al~\cite{qkan}. While their previous work presents a low power figure, it does not include post-synthesis implementation results such as LUT utilization or timing reports. The LUT consumption in this accelerator is determined by software compilation, based on the number of nonlinear functions in KAN models, excluding the hardware consumption of quantization units, adder trees, state machines, and FIFO controls, among others. Therefore, the reported latency and power appear to be based on modeled estimations rather than measured hardware execution. As a result, the practicality of these numbers cannot be fully validated, and the comparison should be interpreted with caution.

In summary, our BiKA accelerator offers the simplest hardware architecture design, the lowest resource consumption, the highest supported clock frequency, and the highest power and hardware efficiency compared to the BNN and QNN accelerators. However, as a trade-off, our BiKA achieves a competitive accuracy on the MNIST dataset, while incurring a higher accuracy loss on CIFAR-10~\cite{Krizhevsky2009}. However, according to our experiment, the refined train strategies and extension of the network model can be expected to improve the accuracy of BiKA on CIFAR-10~\cite{Krizhevsky2009}. Moreover, compared with the extremely high hardware resource consumption shown in~\cite{tran2024KAN} or the complex hardware fabrication required for the previous KAN accelerator designs of Huang et al.~\cite{KAN-Mixed-signal}, our BiKA implementation demonstrates the feasibility of implementing KAN-like networks as hardware accelerators for ultra-lightweight edge devices.

%% file: ISQED2026/conclusion.tex
In this manuscript, we propose a novel multiply-free ultra-lightweight neural network, BiKA, inspired by KAN and BNN. This network highly simplifies the hardware design of accelerators by replacing the multiplier and activation functions with learnable thresholds. Our experiments and implementation results on \emph{Ultra96-V2} show that the BiKA network can reduce the hardware resource consumption by $27.73\%$ and $51.54\%$ compared to BNN and QNN accelerators. 

In our future works, we plan to complete and optimize our BiKA training strategy and explore the potential of BiKA training with larger datasets and complex network models, such as ~\emph{VGG-16}~\cite{Simonyan2014},~\emph{ResNet-50}~\cite{He2015}, \emph{ImageNet}~\cite{imagenet}, etc. Furthermore, to address the major shortcomings in our current design, such as the accuracy loss in the CIFAR-10 datasets, we plan to continually explore the influence of different quantization factors $m$ with larger models.  